# WOODTOUCH, A NEW INTERACTION INTERFACE FOR WOODEN FURNITURE
## DEVELOPMENT OF TOUCH SENSITVE INTERFACES FOR NATURAL INTERACTION WITH WOODEN FURNITURE AND LIGHTING CONTROL

Ivan Arakistain, Mikel Barrado (TECNALIA RESEARCH AND INNOVATION) 2012

## ABSTRACT

**If wood has been with us since time immemorial, being part of our environment, housing and tools, now wood has gain momentum, as it is clear that wood improves our life style.**

**Because of the healthiness, resistance, ecology and comfort, wood is important for all of us, no matter what our life style is.**

**WOODTOUCH Project aims to open a completely new market for furniture and interior design companies, enabling touch interaction between the user and wooden furniture surfaces.**

**Why not switch on or dim the lights touching a wooden table? Why not turn on the heating system? Why not use wood as a touch sensitive surface for domotic control?**

**The furniture designed with this novel technology, offers a wooden outer image and has different touch sensitive areas over the ones the user is able to control all sorts of electric appliances touching over a wooden surface.**

**Keywords: wood, touch, control, sensitive, interaction.**

## INTRODUCTION

Wood is an organic material, a natural composite of cellulose fibers embedded in a matrix of lignin which resists compression. The earth contains about one trillion tones of wood, which grows at a rate of 10 billion tones per year as a renewable carbon absorbing natural material.

Wood has long been used as an artistic medium. It has been used to make sculptures and carvings for millennia. Examples include the totem poles carved by North American indigenous people from conifer trunks, often Western Red Cedar, and the Millennium clock tower, now housed in the National Museum of Scotland in Edinburgh.

There are a lot of touch devices out there, but they lack the warmth and soft touch of natural wood. Why not switch on or dim the lights caressing a wooden table? Why not turn on the heating system? Why not use wood as a touch sensitive surface for domotic control?

WOODTOUCH project opens new business opportunities for wood and furniture industry, particularly related to the integration of lighting systems into furniture.

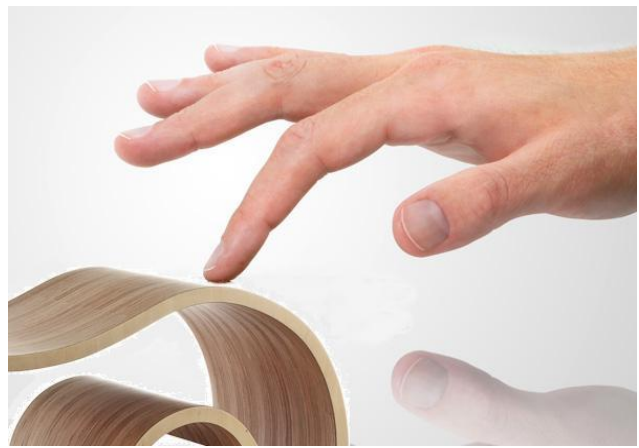

*Figure 1. Wooden furniture with touch interaction capability for a user friendly environment*

## THE CONCEPT

WOODTOUCH is a concept breaking idea supported by HABIC, a cluster of important companies in the Basque Country related to wood industry and interior furnishing design and production evolving almost 100 companies.

The project started in 2010 an it is expected to reach market by 2012 as a purchase option. Prototypes are already available and industrialization is in progress. However, as the project is developing, brand new ideas have come with the need of extra research make WOODTOUCH usable for elderly people, consider mass production cost, profit margin…

*WHY WE TOUCH WOOD*

One of the most charming features of wood is its enormous ability to maintain the warm and of course to transmit it not only in a physical sense, but from a more visual and emotional conception.

Using wood as a constructive and decorative element is to create an ambient full of sensations, to suggest a special and select atmosphere which brigs into memory a full branch of pleasant emotions.

Marketing and merchandising technique experts know it well, that is why it is very frequent to find wooden products in all sorts of commercial stores as well as in the sales points of most prestigious international brands.

Moreover, wood is an excellent acoustic isolator due to its capacity to reduce reverberation time. There is a direct relationship between the increase of wood presence in a room and the reduction of reverberation time. This is why it is a widely used material as an acoustic corrector, for example in sound recording studios. This property of wood can improve our health, as it has been stated that a smaller reverberation time, improves continuous sleeping without interruptions (Berg S, 2001).

What is the origin of the phrase "touch wood" or "knock on wood"? In Pagan belief, people used to believe that evil spirits inhabit in these woods and on knocking the wood the evil inside it will not be able to hear what we just said and presumably will not prevent our hopes from coming true.

The other theory says that the tree and all nature objects were inhabited by God. The touch was meant for luck or protection. The touch was meant to absorb the evil energies.

**Creates subjective comfort.**

Wood helps to create warm and comfortable atmospheres. Due to its natural source, it also contributes to improve the emotional state of inhabitants.

**Dampens impacts.**

Reduces the risk of suffering knee or ligament injuries. It also prevents from suffering wrist or hip breakings in case of falling.

**Is environmentally friendly.**

If we compare the production process of wood made products with the rest of materials, we will see that water and energy consumption is lower in the case of wood. (Over 3 times lower than steel and 2 times lower than concrete). This means that in addition to the saving in its transformation, wood is a clean and ecologic industry because it reduces energy waste and $CO_2$ emissions.

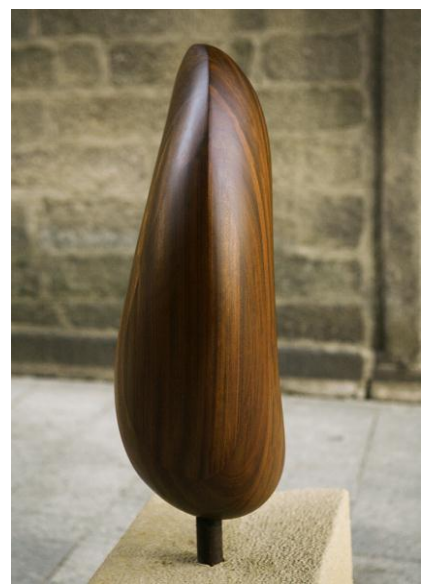

*Figure 2. Wooden sculpture by Jorge Palacios*

The colour, texture and the atmosphere of our environment affects our behaviour that is why we react in a completely different way depending on the sensorial experience we receive from the elements in our environment. We can feel homely, cold, soulless, charming or threatening environments depending on the use of materials, lighting, colours or other perception elements. Fortunately, we feel the environments in which we find wooden elements, as natural, relaxing and quiet.

As a consequence of all this, we can say that consuming wooden products helps to improve our wellbeing and our life quality. These are some of the reasons why wood is called to be the healthy ally.

***WOODTOUCH INTERACTION: "COLD" METAL AND "WARM" WOOD MAY BE THE SAME TEMPERATURE.***

Our hand is not always a good thermometer. When we touch a variety of materials, some will seem warmer or colder than others, even when they are at the same temperature.

The temperature-sensitive nerve endings in the skin detect the difference between inside body temperature and outside skin temperature. When our skin cools down, our temperature sensitive nerves tell us that the object we are touching is cold. An object that feels cold must be colder than our hand, and it must carry our body heat away so that our skin cools down.

Wood and metal are two materials to exemplify this scientific principle. However, they both start at room temperature and are both colder than our hand. They do not feel equally cold because they carry heat away from our hand at different rates.

Wood is an insulator, a very poor conductor of heat. When our hand touches the wood, heat flows from your hand to the wood and warms the wooden surface. Because this heat is not conducted away quickly, the surface of the wood soon becomes as warm as the hand, so little or no additional heat leaves the hand. There is no difference in temperature between the inside of our body and the outside of our skin, so the temperature-sensitive nerves detect no difference in temperature. As a result wood feels warm.

Metal, in contrast, carries heat away quickly. Metal is a good conductor of heat. Heat flows from hand into the metal and then is conducted rapidly away into the bulk of the metal, leaving the metal surface and our skin surface relatively cool. That is why metal feels cool.

Several studies show the preference of wood among users for furnishing, flooring, carpentry and everyday house tools.

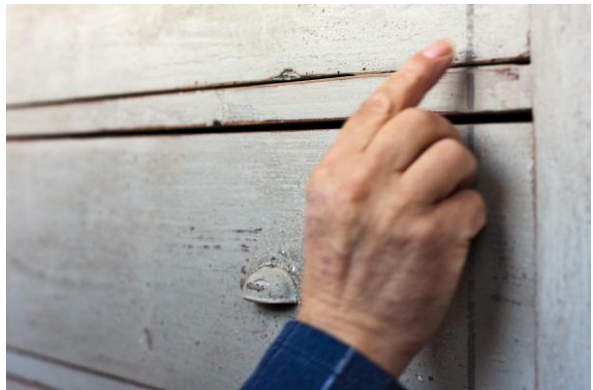

*Figure 3. Touch interaction technology and new uses for furniture*

## TOUCH INTERACTION AND LIGHTING CONTROL

With WOODTOUCH project, "we wanted to capitalize on the tremendous surface area the real world provides".

Touch control is usefull to produce intuitive controls for lighting control, dimming, etc. Some of the applications that were found to be interesting are:

-Access control

-Domotic control

-Accessibility for elderly people

-Energy saving applications to avoid stand by current consumption

-Application in laboratories, operating rooms, etc where disinfection is essential

The rise of touch-based interfaces has revolutionized computing and will probably do the same with ordinary objects, introducing ICT innovations in our everyday life (Escolano F, 2003). Tables for example are an order of magnitude larger than a tablet computer. If we could appropriate these ad hoc surfaces in an on-demand way, we could deliver all of the benefits of mobility while expanding the user's interactive capability.

We are trying to push the boundaries of this rich space of touch and gestures, making gestural interactions available on any surface and with any device.

## FURNITURE AND ENERGY SAVING LIGHTING SYSTEMS

Interior design is undoubtedly linked to lighting, and embedding new distributed lighting systems into the furniture can produce overwhelming environments and increase added value provided by furnishing industry.

Relating the state of technology, it is well worth mentioning that traditional incandescent Light bulbs will disappear progressively of the market in a progressive substitution process towards low consumption products.

The effect of the change in a European level, will mean reducing domestic lighting energy consumption up to an 80%. Moreover, 38 tones less of $CO_2$ will be emitted to the environment, the equivalent to 156 million oil barrels.

From first September 2009 on, the production of the traditional 100 Watts light bulb has stopped in the EU although stores can already sell their stocks. The measure is part of a decision by the European Commision in 2008, with the aim of reducing electric consumption and fume emissions.

Even if the legislation will be progressive until 2016, some efficient lighting solutions are now available and can be integrated with furniture as a whole.

*SOLUTIONS FOR SUSTAINABLE LIGHTING*

LED lamps, manufactured by companies such as Philips,which could replace conventional incandescent and halogen lamps with energy savings up to an 80% and a lifetime of 45,000 hours (Mottier P, 2009), (Fred Schubert E, 2006).

Advantages of LED technology:

-Long use life (50.000 to 100.000 hours) and high luminosity (60 lm/W)

-Low luminous depreciation (30% in 50.000 hours)

-Small footprint

-Dimming possibilities

-Chromatic reproduction index up to 80.

-White light. From 3000K to 6000K

-Low energy consumption

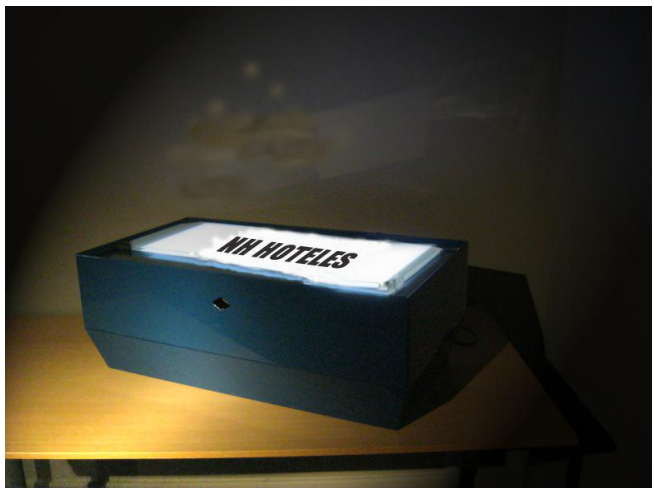

*Figure 4. Electroluminiscence technology for furniture*

LED technology is ideal for Dimming with touch control. The benefits of an LED dimmer switch should be obvious. LED lights last longer, are healthier for the environment and the pocket book because they consume so much less energy, and produce clean, pleasant lighting.

Companies such as Philips will increase their production capability of LED and low consumption fluorescence lighting devices for the foreseen advantages related to energy savings and increasing market demand. The renewing of lamps in an average hotel could suppose a yearly energy saving of 40 Euros for each room simply installing low consumption lamps.

*Integrated compact fluorescent lamps.* In recent years the performance of low consumption (CFLI) energy efficient solutions has increased dramatically. The shape and the design of the latest generation of savers have improved and are easily interchangeable in any fixture.

*Halogen lighting.* Some solutions are designed for maximum light quality and achieve energy savings of up to 50%. They work with all dimmers commonly used to control the light level and create the desired atmosphere.

Another innovative solution is the replacement of fluorescent tubes for high power LED tubes, whose advantages are:

-LED tubes can assume energy savings of 60-80% plus it contains no mercury or lead.

### *NEW SOURCES OF ILLUMINATION*

There are two main ways of producing light: incandescence and luminescence. In incandescence, electric current is passed through a conductor (filament) whose resistance to the passage of current produces heat. The greater the heat of the filament, the more light it produces. Luminescence, in contrast, is the name given to "all forms of visible radiant energy due to causes other than temperature (Wang S, 2000).

There are a number of different types of luminescence, including (among others): electroluminescence, chemiluminescence, cathodoluminescence, triboluminescence, and photoluminescence. Most "glow in the dark" toys take advantage of photoluminescence: light that is produced after exposing a photoluminescent material to intense light. Chemiluminescence is the name given to light that is produced as a result of chemical reactions, such as those that occur in the body of a firefly. Cathodoluminescence is the light given off by a material being bombarded by electrons (as in the phosphors on the faceplate of a cathode ray tube). Electroluminescence is the production of visible light by a substance exposed to an electric field without thermal energy generation.

An electroluminescent (EL) device is similar to a laser in that photons are produced by the return of an excited substance to its ground state, but unlike lasers EL devices require much less energy to operate and do not produce coherent light. EL devices include *light emitting diodes*, which are discrete devices that produce light when a current is applied to a doped p-n junction of a semiconductor, as well as EL displays (ELDs) which are matrix-addressed devices that can be used to display text, graphics, and other computer images. EL is also used in lamps and backlights and it is an ideal complement for furniture to be integrated into flat surfaces, offering a dim lighting, and a wide range of illumination effects controllable with touch.

### CONCLUSSIONS

Touch, voice control, and even gesture control will be coming to daily use devices (Nilsson N, 2001), (Palma J.T, 2003). A common misunderstanding of the natural user interface is that it is somehow mimicry of nature, or that some inputs to a computer are somehow more 'natural' than others. In truth, the goal is to make the user feel like a natural (Russell S, 2004), (Brown J.S, 1983).

Touch control is useful in interior design to produce intuitive controls for lighting, dimming, and electrical appliances while wood is the ideal material for the user friendliness of touch devices.

">PROCEEDINGS DE2012">5

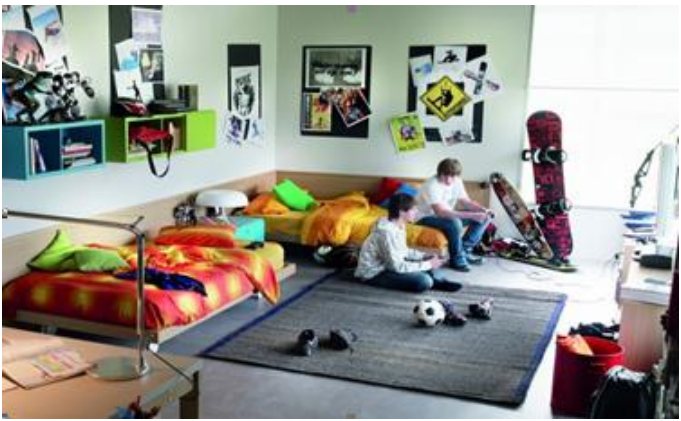

*Figure 5. New design concepts by the companies participating WOODTOUCH project*

*Research Team*

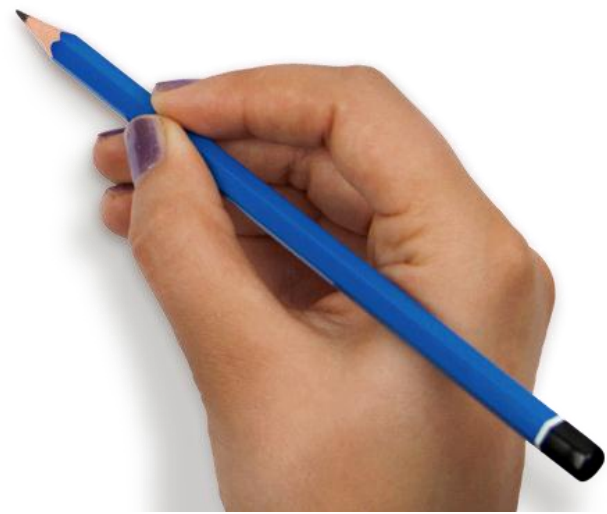